\documentclass[useAMS,usenatbib]{mn2e}
\usepackage{amsmath}
\usepackage{amsbsy}
\usepackage{times}
\usepackage{graphicx}
\usepackage{aas_symbols}
\usepackage{epstopdf}
\usepackage{cases}
\usepackage{color}

\newcommand{\beq}{\begin{equation}}
\newcommand{\eeq}{\end{equation}}
\newcommand{\bea}{\begin{eqnarray}}
\newcommand{\eea}{\end{eqnarray}}
\newcommand{\gae}{\lower 2pt \hbox{$\, \buildrel {\scriptstyle >}\over {\scriptstyle
\sim}\,$}} 
\newcommand{\lae}{\lower 2pt \hbox{$\, \buildrel {\scriptstyle <}\over {\scriptstyle
\sim}\,$}}

\newcommand{\sub}[1]{\ensuremath{_{\rm{#1}}}}
\usepackage{ulem}

\begin{document}

\title[TDE fallback cut-off]{TDE fallback cut-off due to a pre-existing accretion disc}

\author[Kathirgamaraju, Barniol Duran \& Giannios]{Adithan Kathirgamaraju$^{1}$\thanks{Email: akathirg@purdue.edu (AK), rbarniol@purdue.edu (RBD), dgiannio@purdue.edu (DG)}, Rodolfo Barniol Duran$^{1}$\footnotemark[1], Dimitrios Giannios$^{1}$\footnotemark[1] \\
$^{1}$Department of Physics and Astronomy, Purdue University, 525 Northwestern Avenue, West Lafayette, IN 47907, USA 
}

\date{Accepted; Received; in original form ...}

\pubyear{2016}

\maketitle

\begin{abstract}
Numerous tidal disruption event (TDE) candidates originating from galactic centres have been detected (e.g., by {\it Swift} and ASASSN). Some of their host galaxies show typical characteristics of a weak active galactic nucleus (AGN), indicative of a pre-existing accretion disc around the supermassive black hole (SMBH). In this work, we develop an analytic model to study how a pre-existing accretion disc affects a TDE. We assume the density of the disc $\rho\propto R^{-\lambda}$, $R$ being the radial distance from the SMBH and $\lambda$ varying between $0.5 and 1.5$. Interactions between the pre-existing accretion disc and the stream of the tidally disrupted star can stall the stream far from the SMBH, causing a sudden drop in the rate of fallback of gas into the SMBH. These interactions could explain the steep cut-off observed in the light curve of some TDE candidates (e.g., {\it Swift} J1644 and {\it Swift} J2058). With our model, it is possible to use the time of this cut-off to constrain some properties pertaining to the pre-existing accretion disc, such as $\lambda$ and the disc viscosity parameter $\alpha$. We demonstrate this by applying our theory to the TDE candidates {\it Swift} J1644, {\it Swift} J2058 and ASASSN-14li. Our analysis favours a disc profile with $\lambda\sim1$ for viscosity parameters $\alpha\sim0.01-0.1$. 
\end{abstract}

\begin{keywords}
accretion, accretion discs -- methods: analytical -- galaxies: nuclei 
\end{keywords}

\section{Introduction}
Stars orbiting too close to a supermassive blac khole (SMBH) get pulled apart by strong tidal forces, resulting in a tidal disruption event (TDE). In order for this to occur, the pericentre passage of the star must be within the tidal radius $r_{\rm t}=R_{*}(M\sub{bh}/M_*)^\frac{1}{3}$, where $R_*$ and $M_*$ are the radius and mass of the star, respectively, and $M\sub{bh}$ is the mass of the SMBH (\citealp{rees1988}). Following a TDE, about half of the stellar material forms a bound debris stream (BDS) that eventually returns to the vicinity of the SMBH and can lead to tidal disruption flares (\citealp[e.g.,][]{ulmer1999, strubbe2009,lodato2011,guillochon2013}).

Many tidal disruption flare candidates have been detected at multiple wavelengths (a detailed list of these candidates is given in The Open TDE Catalog\footnote{https://tde.space} managed by Katie Auchettl and James Guillochon). There is also evidence that some TDEs  launch powerful relativistic jets (\citealp[e.g.,][]{bloom2011,burrows2011, cenko2012, pasham2015}). Simulations show that magnetic fields from the disrupted material alone may be insufficient to produce a jet in these events (such as in {\it Swift} J1644). A possible solution to this problem is found if the magnetic field of a pre-existing accretion disc is included in these systems (\citealp{tchek2014,kelley2014}).

In recent years, the All-Sky Automated Survey for Supernovae (ASASSN) has detected multiple TDE candidates at galactic centres (\citealp{shappee2014}). In particular, ASASSN-14li (\citealp{holoien2016}; \citealp{brown2016}) shows evidence that its host galaxy harbours a weak active galactic nucleus (AGN; \citealp{alexander2016, prieto2016}). AGNs are powered by an accretion disc which funnels gas into an SMBH. 

Optical observations from a majority of TDE host galaxies yield line emission ratios similar to that of AGNs (\citealp{french2016b}). This supports the claim that TDE rates are enhanced in AGNs (e.g., \citealp{kennedy2016}). Furthermore, many TDE host galaxies show properties similar to post-merger galaxies (\citealp{arcavi2014, french2016a}), and galaxy mergers are believed to be correlated with AGN activity (e.g., \citealp{springel2005}). These results stress the importance of understanding how TDEs behave in the presence of a pre-existing accretion disc.

Previous work by \cite{bonnerot16} investigated the effect of TDEs in the presence of a very low-density, static corona. They found that the interaction between the BDS and the corona leads to Kelvin-Helmholtz instabilities, which may, in some cases, dissolve the TDE stellar stream. Their results were relevant for the disruption of giant stars. In this work, we study how a rotationally supported pre-existing accretion disc can affect the dynamics of the BDS. Our study identifies observable effects even in the case of the disruption of main sequence stars. In general, the accretion disc can alter the fallback of the BDS on to the SMBH, thereby modifying the duration of the flare and their observed light curves. We find our results to be very sensitive to the accretion disc model. This enables us to use TDEs to probe the structure of accretion discs at galactic centres.

In Section 2, we describe the model used for the pre-existing accretion disc and the BDS. In Section 3, we calculate the dynamics of the BDS as it travels through the accretion disc and study how this might affect TDE observations. In Section 4, we present our results and apply our theory to TDE candidates around accreting SMBHs in Section 5. We discuss our results and conclude in Section 6.

\section{Modelling the pre-existing accretion disc and the bound debris stream}

In this work, we consider an SMBH surrounded by a pre-existing accretion disc. The disc extends out to a scale comparable to the Bondi radius, typically a fraction of a parsec. Before the TDE, as the star gets closer to the black hole, it travels unimpededly towards it. The stellar density is much larger than the density of the accretion disc, so the stellar trajectory is not affected by the presence of the accretion disc. As the star is disrupted, the bound stellar material, now in the form of a BDS, expands as it traverses through the accretion disc. We assume a geometrically thick disc, as expected in low-luminosity AGN systems. We take the height of the disc ($H$) to be on the order of $R/2$, $R$ being the radial distance from the SMBH. Using the solid angle subtended by the disc, we estimate roughly half of all possible stellar orbits will lie within the plane of the disc in our model. For simplicity, in what follows, we assume a coplanar disc-BDS configuration.

\subsection{The pre-existing accretion disc around the SMBH}

We parametrize the density of the pre-existing accretion disc around the black hole by assuming it accretes at some fraction of the Eddington accretion rate. The Eddington luminosity of a black hole of mass $M\sub{bh}$ is $L_{\rm edd} \approx 10^{44} M_6$ erg/s, where $M_6=M_{\rm bh}/10^6M_{\odot}$.  The mass accretion rate that corresponds to this luminosity depends on the efficiency $\eta$, and is given by
\beq
\dot{M}_{\rm edd} = \frac{L_{\rm edd}}{\eta c^2} = \frac{4\pi GM\sub{bh}m\sub{p}}{\eta\sigma\sub{T}c},
\eeq
where $\sigma\sub{T}$ is the Thomson cross-section for an electron, $m\sub{p}$ is the mass of a proton, $c$ is the speed of light and $G$ is the gravitational constant.
We can normalize a given mass accretion rate to the black hole, $\dot{M}$, to this quantity by defining $\dot{m} = \dot{M}/{\dot{M}_{\rm edd}}$. Therefore, if a pre-existing accretion disc is present, its mass accretion rate is given by
\beq
\dot{M} =\frac{4\pi GM\sub{bh}m\sub{p}}{\eta\sigma\sub{T}c}\dot{m}=(10^{23} \, {\rm g/s}) \, \dot{m} M_6 \eta^{-1}.
\label{expected_mdot}
\eeq

The mass accretion rate from a thick disc can be obtained by extending the \cite{shakura1973} thin disc model to a scale height of the disc $H\sim R/2$, where $R$ is the radial distance from the black hole, and it is given by
\beq
\dot{M} = \frac{\pi}{2} R^2 \alpha v_{\rm k} \rho,
\label{mdot_theory}
\eeq
where $\alpha$ is the viscosity parameter, $v_{\rm k}$ is the Keplerian speed and $\rho$ is the density of the disc at a distance $R$.  The Keplerian speed at a distance $R$ from the black hole is $v_{\rm k} = c \sqrt{R_{\rm g}/R}$, where $R\sub{g} = G M\sub{bh} / c^2 \approx 1.5 \times 10^{11} M_6 \,{\rm cm} $.

The number density of particles in the disc is $n = \rho / m\sub{p}$. At a distance $R_0 = 10 R\sub{g}$, the number density $n_0$ can be obtained using equations (\ref{expected_mdot}) and (\ref{mdot_theory}), and is given by
\beq
n_0=\frac{8c^2}{10^{\frac{3}{2}}\sigma\sub{T}G}\left(\frac{\dot{m}}{\alpha\eta}\right)M\sub{bh}^{-1} \approx (1.7 \times 10^{12} \, {\rm cm^{-3}}) \, \dot{m} M_6^{-1} \alpha^{-1} \eta^{-1}.
\eeq
From now on, we will use $\mu=\frac{\alpha\eta}{\dot{m}}$ to parametrize the disc density close to the SMBH. We will explore typical values of $\eta,\alpha\lesssim 0.1$ and $\dot{m}\lesssim 0.01$ for which $\mu$ may be of order unity .
We assume that far from the SMBH ($R\gtrsim R_0$), the density of the accretion disk drops as $\propto R^{-\lambda}$. Therefore, the density at distance $R$ is 
\beq
n = n_0 \left(\frac{R}{R_0}\right)^{-\lambda}.
\eeq
Throughout this work, we consider $\lambda$ to be within $0.5-1.5$, which are the limits set by the Convection-Dominated Accretion Flow (CDAF; \citealp{narayan2000,quataert2000} and Advection-Dominated Accretion Flow (ADAF; \citealp{narayan1994}) models for accretion discs, respectively. We will show how TDE observations might allow us to constrain
the important parameter $\lambda$, which sets the density profile of the disc.

\subsection{Geometry and dynamics of the bound debris stream} \label{sec:dynamics}

We assume the star approaches the SMBH on a highly elliptical orbit with a pericentre distance $r\sub{p}\sim r\sub{t}$, in order to avoid the complications that arise from deeply penetrating orbits. After the star is disrupted by the SMBH, the stellar debris forms an elongated, cylindrical stream (\citealp{guillochon15}) of diameter $h$. The stream initially evolves under self-gravity, with its diameter scaling as $h\propto R^{1/4}$, where $R$ here is the distance from the SMBH to the stream (\citealp{kochanek1994}; \citealp{guillochon2014}). Initially, the temperature of the stream evolves adiabatically as $T\propto\rho^{(\gamma-1)}$, where $\gamma=5/3$ and $\rho$ is the density of the stream (\citealp{kasen2010}). When the stream has cooled to a temperature  $T\sub{rec}\approx 10^4$K it undergoes recombination, during which the temperature of the stream (for a solar-type star) remains roughly constant (\citealp{kasen2010}). When the stream starts recombining it no longer self-gravitates, instead it expands laterally at its internal sound speed $c\sub{s}\approx\sqrt{\gamma k\sub{b} T\sub{rec}/m\sub{p}} \approx 1.6\times 10^6$ cm/s (\citealp{guillochon15}).  After recombination, the stream continues to evolve adiabatically.  Following this evolution, the diameter $h$ of the BDS is given in equation (10) of  \cite{guillochon15}  as
\beq
h=2R_*\bigg(\frac{{\rm min}[R,r_{\rm rec}]}{r\sub{p}}\bigg)^{\frac{1}{4}} + {\rm max}[c\sub{s}(t-t_{\rm rec}),0],
\label{h}
\eeq
where $r_{\rm rec}$ and $t_{\rm rec}$ are the radius and time at which recombination occurs and $r\sub{p}$ is the pericentre distance of the star. Throughout this work, we will assume $r\sub{p}=r\sub{t}$. We can relate the apocentre distance of each stream element to the time of its apocentre passage using Kepler's third law
\beq
r\sub{apo}=2a-r\sub{t}\approx 2a=2\bigg(\frac{GM\sub{bh}\,t\sub{apo}^2}{\pi^2}\bigg)^{\frac{1}{3}},
\label{r}
\eeq 
where $a$ is the semi-major axis and we have used the fact that $r\sub{t}\ll a$. Unless otherwise specified, we take $t=t\sub{apo}$ and $r=r\sub{apo}$ from now on. According to equation (\ref{h}), for times $t\gg t_{\rm rec}$, $h\approx c\sub{s}t$ is a good approximation \citep{guillochon15}. 

A useful time-scale is the fallback time of the most bound material ($t\sub{fall}$). The specific orbital energy of the most bound material is $\sim GM\sub{bh}R_*/r\sub{t}^2$  (e.g., \citealp{lacy1982}). Using Kepler's law in terms of orbital energy, we find
\beq
t\sub{fall} = \frac{\pi R_*^{\frac{3}{2}}}{M_*}\bigg(\frac{M\sub{bh}}{2G}\bigg)^{\frac{1}{2}}\approx (3.6\times 10^6 s)\frac{r_*^{\frac{3}{2}}M_6^{\frac{1}{2}}}{m_*}
\label{tfall},
\eeq
where $m_*=M_*/M_{\odot}$ and $r_*=R_*/R_{\odot}$.

We approximate the length of the stream to be $l\approx 2a\approx r$. Assuming a uniform stream, we can calculate its number density as
\beq
n_{\rm s} = \frac{M_*}{2\pi (h/2)^2rm\sub{p}}.
\label{debris_density}
\eeq
The factor of 2 in the denominator appears because roughly half of the star remains bound to the SMBH.
Another quantity we will use later is the apocentre velocity of the stream element. It is given by the vis-viva equation
\beq
v\sub{apo}=\sqrt{GM\sub{bh}\bigg(\frac{2}{r}-\frac{1}{a}\bigg)}.
\eeq
We use the exact expression for $r$ given in equation (\ref{r}) and Taylor-expand it in powers of $r\sub{t}/a$. Keeping only the first-order term (since $r\sub{t}\ll a$), we find 
\beq
v\sub{apo}\approx\frac{1}{a}\sqrt{GM\sub{bh}\frac{r\sub{t}}{2}}.
\eeq

\section{Interactions between the pre-existing disc and the BDS} \label{interaction}

All interactions between the disc and the BDS are considered at the apocentre passage of each stream element. Each stream element travels the slowest at its apocentre passage, therefore the time of interactions with the disc is maximized at this point. Also the density ratio of the stream to that of the disc is lower at this point. This means that the disc can have a more significant impact on the BDS at or after apocentre passage. 

We can approximate the relative velocity between the BDS and the disc at apocentre to be the Keplerian velocity (the assumed velocity of the accretion disc), since 
\beq
\frac{v\sub{apo}}{v\sub{k}(r)}\approx\sqrt{\frac{r\sub{t}}{a}}\ll1.
\eeq
Therefore, at apocentre, we can assume the BDS is stationary and the disc material ``rams" the stream with speed $\sim v\sub{k}$. Since this paper focuses on interactions at apocentre passage, our results are insensitive to whether the disc is prograde or retrograde to the BDS as the interactions at apocentre will be described similarly in both cases.

Interactions between the disc and the BDS can manifest in the form of a shock which propagates through the BDS. These shocks can directly affect the trajectory of the BDS and prevent it from reaching the black hole (Section \ref{shockinteraction}). The velocity difference between the disc and the shocked BDS can lead to the development of Kelvin--Helmholtz (KH) instabilities (Section \ref{KHinstability}). If this instability grows sufficiently fast within the BDS, it can completely disrupt it and also prevent the BDS from reaching the black hole. In some cases, the interactions may not be strong enough to form a shock or to trigger the instabilities. For these cases, we estimate the momentum imparted by the disc on to the BDS and assess whether the BDS trajectory is perturbed enough to prevent it from reaching the black hole (Section \ref{Momentum}). 

\subsection{Interaction via shocks} \label{shockinteraction}

The interaction between the disc and the BDS produces a shock that travels laterally through the stream. The speed of the shock front, in the frame of the shocked fluid, can be approximated by equating the ram pressure of the disc with the ram pressure of the shock.  This yields 
\beq
v_{\rm sh} \approx \frac{1}{3} \left( \frac{n}{n\sub{s}} \right)^{\frac{1}{2}} v_{\rm k}=\frac{10^{\frac{1}{4}}c}{15}\left(\frac{\pi m\sub{p}}{\mu\sigma\sub{T}M_*}\right)^{\frac{1}{2}}h\left(\frac{10GM\sub{bh}}{c^2r}\right)^{\frac{\lambda}{2}}.
\label{vsh}
\eeq

The time taken for the shock to cross the debris stream at apocentre is
\beq
t_{\rm cross} = \frac{h}{4v_{\rm sh}}=  \frac{3\times 5^{\frac{3}{4}}}{2^{\frac{9}{4}}c}\left(\frac{\mu\sigma\sub{T}M_*}{\pi m\sub{p}}\right)^{\frac{1}{2}}\left(\frac{c^3t\sub{apo}}{\pi\,5^{\frac{3}{2}} GM\sub{bh}}\right)^{\frac{\lambda}{3}},
\eeq
which, given the geometry of the problem, turns out to be independent of $h$. The factor of 4 appears in the denominator due to the reference frame considered: $4 v\sub{sh}$ corresponds to the speed at which the shock propagates into the stream, in the stream rest frame.  These calculations are valid only for a strong shock (i.e. when the Mach number, $v\sub{sh}/c\sub{sound}\gg1$). 

The time when $t\sub{cross}\approx t\sub{apo}$ yields the time when the shock completely crosses the BDS at apocentre (``shock-crossing time"), which is given by
\beq
t\sub{shock}=\Bigg[ \frac{3\times 5^{\frac{3}{4}}}{2^{\frac{9}{4}}c}\left(\frac{\mu\sigma\sub{T}M_*}{\pi m\sub{p}}\right)^{\frac{1}{2}}\left(\frac{\pi\,5^{\frac{3}{2}}GM\sub{bh}}{c^3}\right)^{-\frac{\lambda}{3}}\Bigg]^{\frac{3}{3-\lambda}}.
\label{tshock}
\eeq 
Considering the range of values for the parameter $\lambda = 0.5 - 1.5$, the shock-crossing 
time for the fiducial parameters is 
\begin{subnumcases}{t\sub{shock}\approx }
(6\times 10^6 {\rm s})\,\mu^{\frac{3}{5}} m_*^{\frac{3}{5}} M_6^{-\frac{1}{5}}& $\lambda=$0.5\\
(10^8 {\rm s})\,\mu^{\frac{3}{4}} m_*^{\frac{3}{4}}M_6^{-\frac{1}{2}}& $\lambda=$1\\
(7\times 10^{9} {\rm s})\,\mu m_*M_6^{-1}& $\lambda=$1.5.
\end{subnumcases}
Since the density of the disc $\rho\propto R^{-\lambda}$, larger values of $\lambda$ imply a steeper drop in density. This results in weaker interactions, which lead to a lower shock velocity and a longer time for the shock to cross the BDS. However, for flat disc profiles ($\lambda\sim0.5$) and $\mu\sim1$, the shock crosses within a few months after the TDE.

The shock alters the velocity of the BDS, but if the shock is too weak, it may not sufficiently modify the trajectory of the BDS so as to prevent the stream from falling back to the SMBH. In these cases, we find that the KH instability that grows in the disc--BDS interface is able to stall the BDS. We show below that as long as a shock exists,  KH instabilities develop along the interface of the disc and the BDS. These instabilities will disrupt the entire stream on a time-scale similar to the shock-crossing time (at $t\sub{shock}$), preventing the BDS from falling back to the black hole.

\subsection{Kelvin--Helmholtz instability}\label{KHinstability}

As the shock crosses the BDS, the shocked fluid becomes prone to KH instability (since it is in pressure balance with the disc material). This instability will disrupt the stream when it grows to a wavelength of size $\sim h$ (\citealp{bonnerot16}). The time taken for this growth is given by (e.g. \citealp{choudhuri1998})
\beq
t\sub{KH}\approx \frac{h}{v\sub{k}}\left(1+\frac{n\sub{s}}{n}\right)^{\frac{1}{2}}=\frac{h}{3v\sub{sh}}\left(1+\frac{n}{n\sub{s}}\right)^{\frac{1}{2}},
\eeq
where we have used equation (\ref{vsh}) to obtain the last equality. In deriving the above equation, we ignore self-gravity of the stream since $t\sub{rec}\textless t\sub{fall}$ in all cases considered throughout this paper (see Section \ref{sec:dynamics}). For the parameters considered in this work, the disc is less dense than the stream (i.e. $n/n\sub{s} \ll 1$), allowing us to approximate $t\sub{KH}\approx h/3v\sub{sh}\approx t\sub{cross}$. This approximation is valid even if the shock is weak. Therefore, as long as a shock exists, the KH instability will disrupt the BDS when the shock completely crosses the stream at $t\sub{shock}$. So the fallback of material on to the SMBH gets cut-off at $t\sub{shock}$ provided there is a shock. A shock exists when 
\beq
\frac{T\sub{sh}}{T\sub{s}}(t\sub{shock})\ge 1,
\label{shock}
\eeq
where $T\sub{sh}$ is the temperature of the shocked fluid and $T\sub{s}$ is the temperature of the pre-shocked BDS, and we have evaluated the ratio at $t\sub{shock}$. For a more detailed calculation on the shock conditions see the Appendix.

\subsection{Momentum imparted by the disc}\label{Momentum}

In some situations, the interaction between the disc and the BDS is not strong enough to produce a shock (e.g., if the disc density is too low or if the temperature of the BDS is too high). For these cases condition (\ref{shock}) is not satisfied, so a shock does not propagate through the BDS. When there is no shock, the momentum imparted by the disc on to the BDS might be sufficient to affect its trajectory. Consider a stream element at apocentre, its initial momentum will be $v\sub{apo}(t)dm$, where $t$ corresponds to the apocentre time of this element and $dm=\frac{M_*}{2r}dr$ (we take a slice of the stream in the radial direction). Let $\rho$ be the density of the disc. The disc 'sees' a cross-section $h(t)dr$ of the stream. The amount of momentum transferred from the disc on to the stream element after some time $\tau$ since disruption will be $\sim\rho v\sub{k}^2h\tau dr$. From momentum conservation, we obtain
\beq
v\sub{s}(\tau)dm=v\sub{apo}(t)dm-\rho v\sub{k}^2h\tau dr,
\eeq
where $v\sub{s}$ is the final speed of the stream element. In order to stall the stream, that is, to prevent it from reaching the black hole, we will assume the momentum imparted by the disc should reduce the BDS velocity by at least a factor of $\sim 2$ (i.e. $v\sub{s}(\tau)=0.5v\sub{apo}$). Expressing the above equation in terms of time, we find that the time for this velocity change to occur is 
\beq
\tau=\frac{5^{\frac{3}{2}}\pi^{\frac{2}{3}}G^{\frac{1}{6}}\sigma\sub{T}\,\mu}{16\,c\sub{s}\,c^2m\sub{p}}\,M_*^{\frac{5}{6}}R_*^{\frac{1}{2}}M\sub{bh}^{\frac{1}{3}}\left(\frac{c^3}{5^{\frac{3}{2}}\pi GM\sub{bh}}\right)^{\frac{2\lambda}{3}}t^{\frac{2\lambda-5}{3}},
\eeq
where we used $h=c\sub{s}t$, which is valid since $t\sub{rec}\textless t\sub{fall}$.

The maximum time each stream element has to interact with the disc is of the order of its apocentre time $t$. Equating $\tau\approx t$  yields the 'stalling time' of the stream (due to impulse from the disc), given by
\beq
\tau\sub{stall}= \Bigg(\frac{5^{\frac{3}{2}}\pi^{\frac{2}{3}}G^{\frac{1}{6}}\sigma\sub{T}\,\mu}{16\,c\sub{s}\,c^2m\sub{p}}\,M_*^{\frac{5}{6}}R_*^{\frac{1}{2}}M\sub{bh}^{\frac{1}{3}}\left(\frac{c^3}{5^{\frac{3}{2}}\pi GM\sub{bh}}\right)^{\frac{2\lambda}{3}}\Bigg)^{\frac{3}{8-2\lambda}}.
\label{tstall}
\eeq
Considering the range of values for the parameter $\lambda = 0.5$--$1.5$, the stalling time of the stream for the fiducial parameters is
\begin{subnumcases}{\tau\sub{stall}\approx }
(2\times 10^7 {\rm s})\,\mu^{\frac{3}{7}}\,r_*^{\frac{3}{14}}m_*^{\frac{5}{14}}& $\lambda=$0.5\\
(10^8 {\rm s})\,\mu^{\frac{1}{2}}\,r_*^{\frac{1}{4}}m_*^{\frac{5}{12}}M_6^{-\frac{1}{6}}& If $\lambda=$1\\
 (2\times 10^9 {\rm s})\,\mu^{\frac{3}{5}}\, r_*^{\frac{3}{10}}m_*^{\frac{1}{2}} M_6^{-\frac{2}{5}}& $\lambda=$1.5.
\end{subnumcases}
As in the case of the shock-crossing time, the stalling time increases with $\lambda$ since the density of the disc falls off more steeply as $\lambda$ increases. 

\subsection{The cut-off time}\label{sec:cutofftime}

From the above calculations, we can estimate a time-scale at which the fallback of the BDS on to the SMBH ceases. We will call this the 'cut-off time' ($t\sub{cutoff}$) and define it as 

\begin{subnumcases}{t\sub{cutoff}=}
t\sub{shock}-t\sub{fall}& if $\frac{T\sub{sh}}{T\sub{s}}(t\sub{shock})\ge 1$ \label{shockstall}\\ 
\tau\sub{stall}-t\sub{fall}& otherwise. \label{mstall}
\label{cutoff}
\end{subnumcases}

\begin{figure}
\includegraphics[width=9cm]{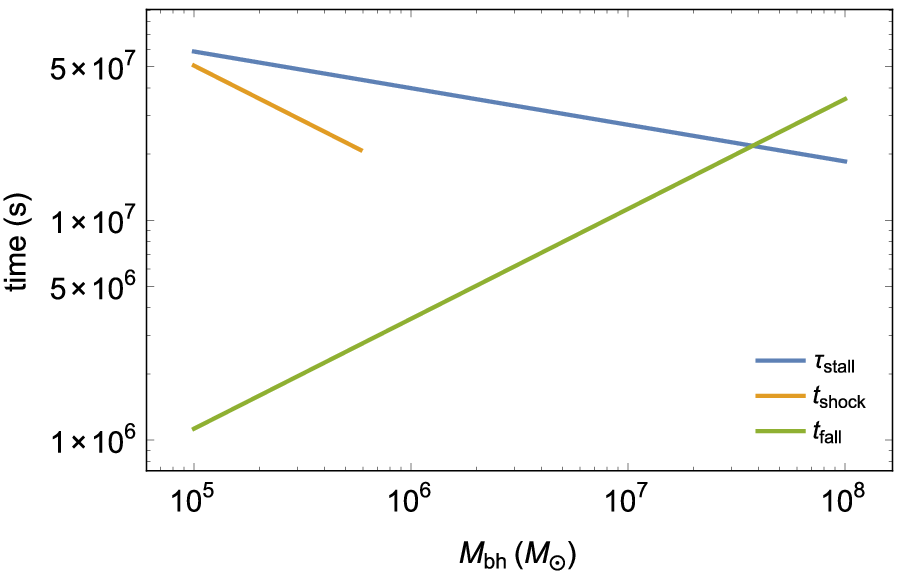}\\
\includegraphics[width=9cm]{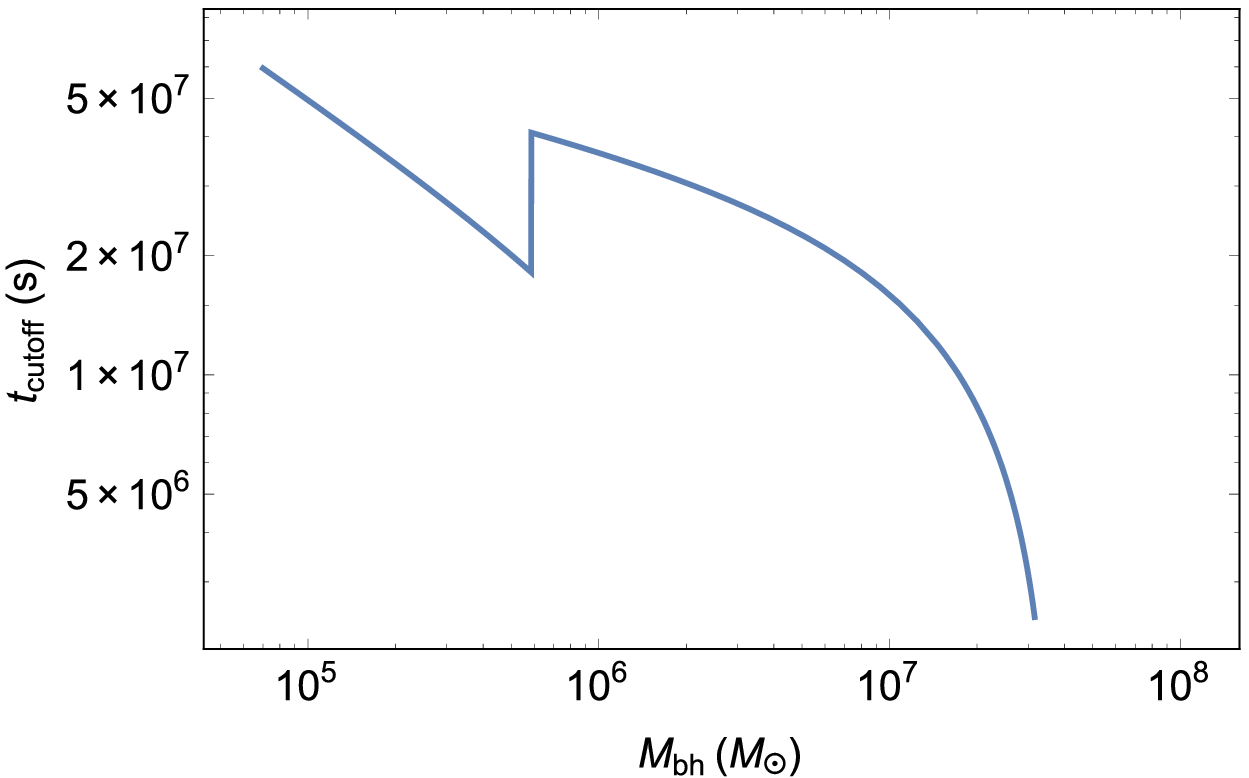} 
\caption{Various time-scales (top panel) and the cut-off time (bottom panel) associated with a TDE that interacts with a pre-existing accretion disc. We assume a solar-type star and take $\lambda=1, \dot{m}=\alpha=0.01, \eta=0.1$ (see equations (\ref{tfall}), (\ref{tshock}), (\ref{tstall}) and  (23) for analytic expressions of time-scales). The discontinuous jump in the cut-off time at $5\times 10^5 M_{\odot}$ appears because the stalling mechanism of the BDS changes from being due to a shock crossing the BDS to the disc imparting its momentum to the BDS.}
\label{fig1}
\end{figure}

In order to represent observations more accurately, we assume the TDE is detected at a time similar to the fallback time of the most bound material, hence we subtract $t\sub{fall}$ in the above equation. We show a plot of $t\sub{shock},\tau\sub{stall}$ and $t\sub{fall}$ versus SMBH mass in Fig. \ref{fig1} (top panel), in the same figure we show the corresponding cut-off time $t\sub{cutoff}$ (bottom panel). Since these time-scales depend on both the disc density and its profile, as an example for this plot we choose $\lambda=1$, $\dot{m} = \alpha = 0.01$ and $\eta=0.1$ (other parameters are considered and discussed in Section \ref{results}). We find that for lower mass SMBHs, the cutoff time is due to shock-crossing given by equation (\ref{shockstall}), but for SMBHs of mass $\gtrsim 5\times 10^5 M_{\odot}$, there is no shock in the BDS since condition (\ref{shock}) is not satisfied. In this case the stalling mechanism transitions to impulse caused by the disc; see equation (\ref{mstall}). This transition is seen as a discontinuous jump in the cut-off time.
 
%%%%%%%%%%%%%%%%%%%%%%%%%%%%%%%%%%%
%%%%%%%%%%%%%%%%%%%%%%%%%%%%%%%%%%%

\subsection{Identifying cut-off time from observations}\label{luminosityratio}

Tidal disruption flares caused by the BDS are interpreted in two ways: (1) they are powered by the accretion of the bound material on to the SMBH, which can manifest in the form of a jet or outflows (\citealp[e.g.,][]{strubbe2009, giannios2011,bloom2011}), (2) they are caused by emission from the resulting accretion disc, which forms when the BDS that shocks itself circularizes (\citealp[e.g.,][]{cannizzo1990,ulmer1999,guillochon2014}). In both these scenarios, the flare should drop in luminosity (or cease) when the mass feeding the flare stops. The time at which this happens should coincide with the cut-off time, assuming the viscosity and the radiative cooling time-scales of the system powering the flare are shorter than $t\sub{cutoff}$. Hence, regardless of which mechanism is powering the flare, the cut-off time is a robust observable that can be used in our model to determine the properties of the pre-existing accretion disc.

\begin{figure}
\includegraphics[width=9cm]{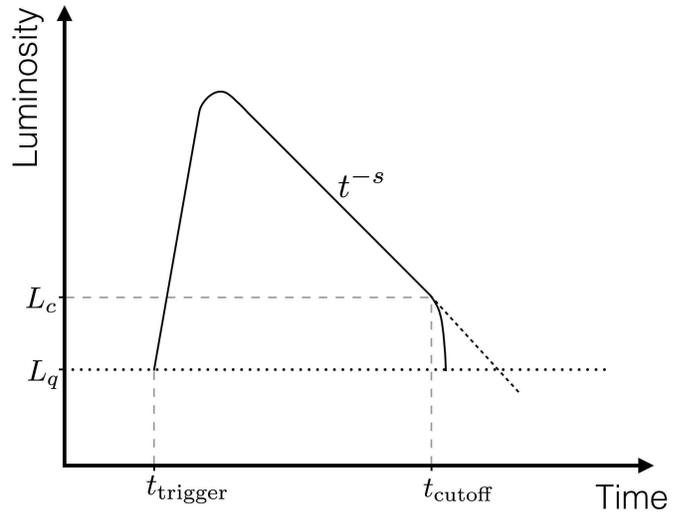}
\caption{Diagram illustrating the light curve of a tidal disruption flare. At $t\sub{c}$ we expect the luminosity to sharply drop from $L\sub{c}$ to the quiescent value $L\sub{q}$ (horizontal dotted line) due to the interaction of the BDS and a pre-existing accretion disc. The dashed line indicates what the flare would look like if no pre-existing accretion disc is present.}
\label{diagram}
\end{figure}

After the cut-off time we might observe a steep decay in the luminosity of the flare since the stream of mass powering the flare has ceased. If the host galaxy harbours an AGN, the tidal disruption flare luminosity will decay to some constant quiescent value $L\sub{q}=\eta\dot{M}c^2$ corresponding to the AGN power. Hence from observations, we can identify the cut-off time when the light curve of the flare becomes significantly steeper. The steeper the decay, the easier it will be to identify $t\sub{cutoff}$. We can gauge the steepness of this decay by calculating the ratio of the flare luminosity at the cut-off time ($L\sub{c}$) to the quiescent luminosity $L\sub{q}$ of the AGN (see Fig. \ref{diagram} for an illustration of the cut-off time and associated luminosities). We can estimate $L\sub{c}$ by assuming the mass of the BDS falls back at a rate $\dot{M}\sub{fb}\propto t^{-s}$
\beq
\begin{split}
L\sub{c}=&f\sub{b}\,\eta\sub{fb}\dot{M}\sub{fb}(t\sub{cutoff})c^2=(s-1)f\sub{b}\,\eta\sub{fb}\frac{M_*c^2}{2\,t\sub{fall}}\left(\frac{t\sub{cutoff}}{t\sub{fall}}\right)^{-s},
\label{Lc}
\end{split}
\eeq
where $f\sub{b}$ is the beaming factor of the flare ($f\sub{b}=1$ if the flare is not beamed and $\textgreater 1$ if the flare is beamed), $\eta\sub{fb}$ is the radiative efficiency from the accretion of the fallback material. From this we can calculate $L\sub{c}/L\sub{q}$. If this ratio is large, it implies a substantial steepening in luminosity at $t\sub{cutoff}$. This in turn would make identifying $t\sub{cutoff}$ easier.

%%%%%%%%%%%%%%%%%%%%%%%%%%%%%%%%%%%
%%%%%%%%%%%%%%%%%%%%%%%%%%%%%%%%%%%

\begin{figure}
\includegraphics[width=9cm]{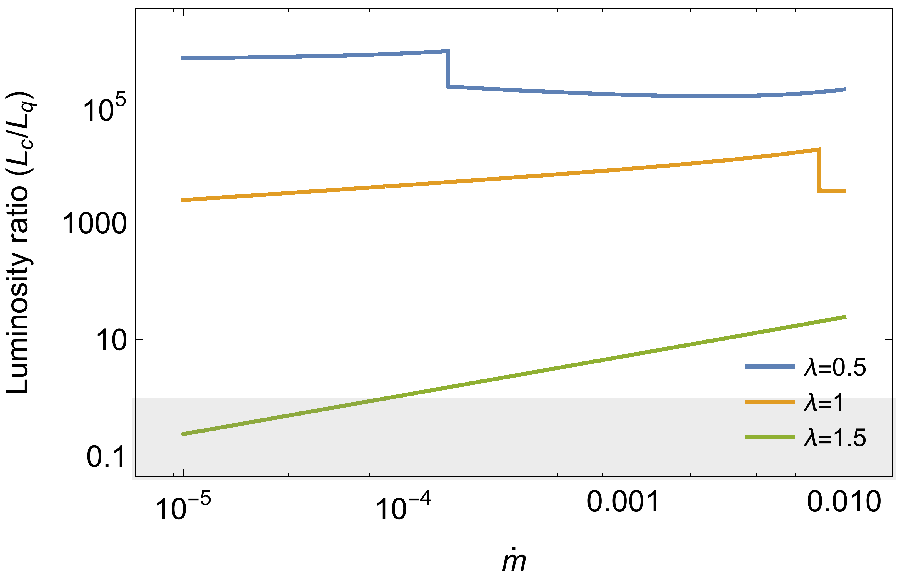}\\
\includegraphics[width=9cm]{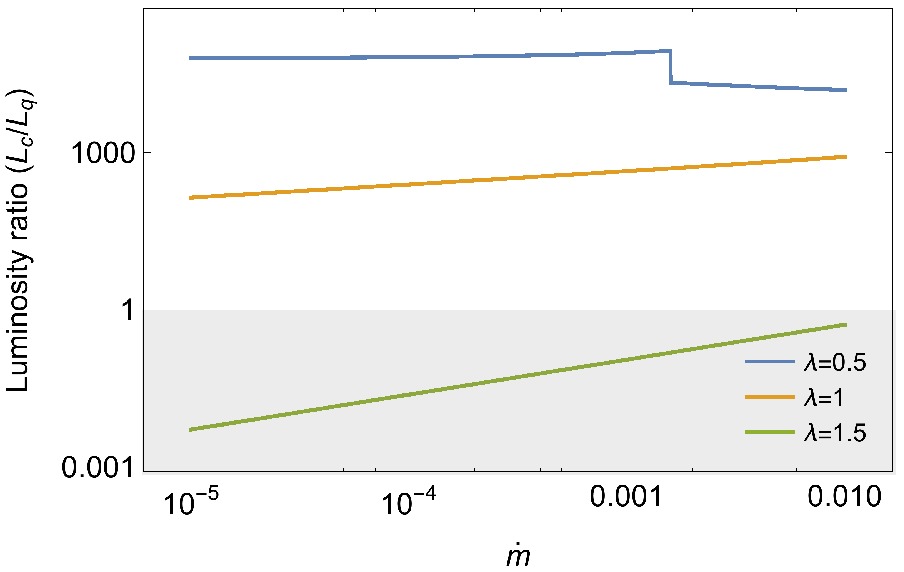}
\caption{ $L\sub{c}/L\sub{q}$ versus accretion rate $\dot{m}$ (normalized to Eddington) for SMBH mass $10^6 M_{\odot}$ and multiple values of $\lambda$ with $\alpha=0.01$ (top panel) and $\alpha=0.1$ (bottom panel). We assume a solar-type star and $s=5/3$, $\eta=0.1$, $f\sub{b}\eta\sub{fb}=1$. In most cases this ratio is very large, which suggests a high chance of identifying $t\sub{cutoff}$, the shaded area shows the region in which $t\sub{cutoff}$ cannot be identified.}
\label{lratio}
\end{figure}

Theory predicts the temporal index, $s$, should be $s=5/3$ (\citealp{rees1988,phinney1989}), and in Fig. \ref{lratio} we plot the luminosity ratio $L\sub{c}/L\sub{q}$ versus $\dot{m}$ using $s=5/3$ for multiple values of $\alpha$ and $\lambda$ keeping the SMBH mass fixed at $10^6 M_{\odot}$. Here we use the analytic expression for $t\sub{cutoff}$ in $L\sub{c}$; see equation (23). We see that in most cases, $L\sub{c}/L\sub{q}\gg1$, which means $t\sub{cutoff}$ should be easily observable (provided $L\sub{c}$ is above the sensitivity limits of the instrument). In Fig. \ref{lratio}, we see the luminosity ratio can be less than unity for larger values of $\lambda \approx 1.5$, which means $t\sub{cutoff}$ cannot be identified in these cases, since the disc density is small and the interaction between the disc and the BDS is weak. We show in the next section that the cut-off time for these large values of $\lambda$ is on the order of hundreds of years and therefore will be of little observational relevance.
 
%%%%%%%%%%%%%%%%%%%%%%%%%%%%%%%%%%%
%%%%%%%%%%%%%%%%%%%%%%%%%%%%%%%%%%%

\section{Results}\label{results}

\begin{figure}
\includegraphics[width=9cm]{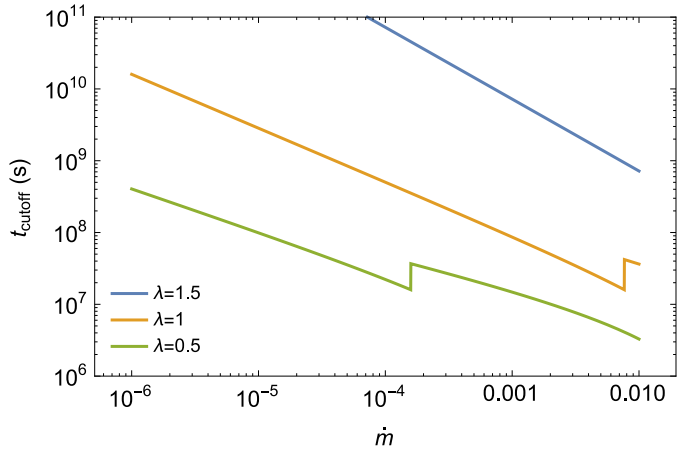}
\caption{Cut-off time versus accretion rate $\dot{m}$ (normalized to Eddington) for a fixed SMBH with amass of $10^6M_{\odot}$ and multiple density profiles ($\lambda$) of the disc. We assume a solar-type star and take $\alpha= 0.01$ and $\eta=0.1$ (see Section \ref{interaction} for analytic expressions of $t\sub{cutoff}$). For disc profiles of $\lambda\lesssim1$ and reasonably high accretion rates $\dot{m}\gtrsim 10^{-4}$, the cut-off time is $\sim 1$ year after the TDE, and can therefore be observable.}
\label{fig2}
\end{figure}

In Fig. \ref{fig2}, we show the cut-off time versus accretion rate ($\dot{m}$) for an SMBH of $10^6 M_{\odot}$ for multiple density profiles of the disc. We see the cut-off time differs significantly for different values of $\lambda$, indicating that this time-scale is very sensitive to the density profile of the disc. 

\begin{figure}
\includegraphics[width=9cm]{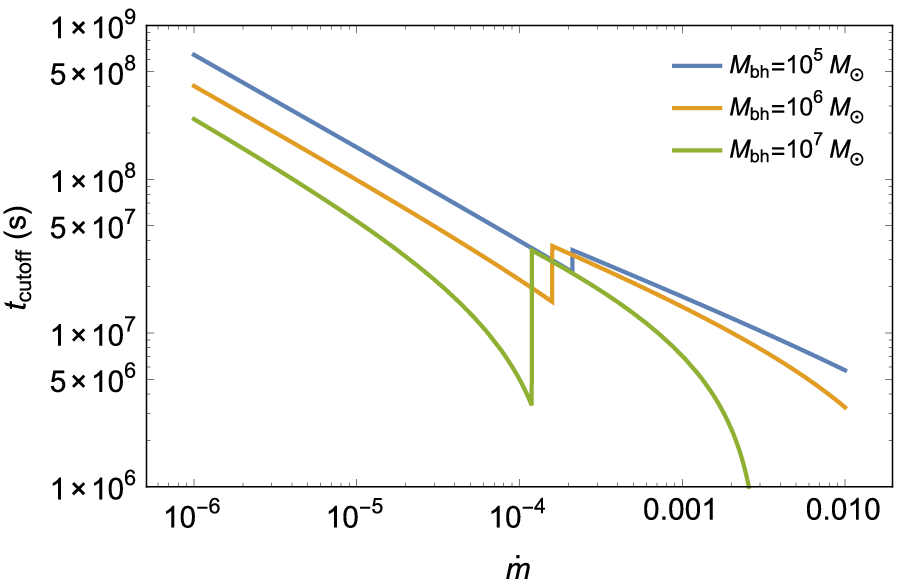}\\ \includegraphics[width=9cm]{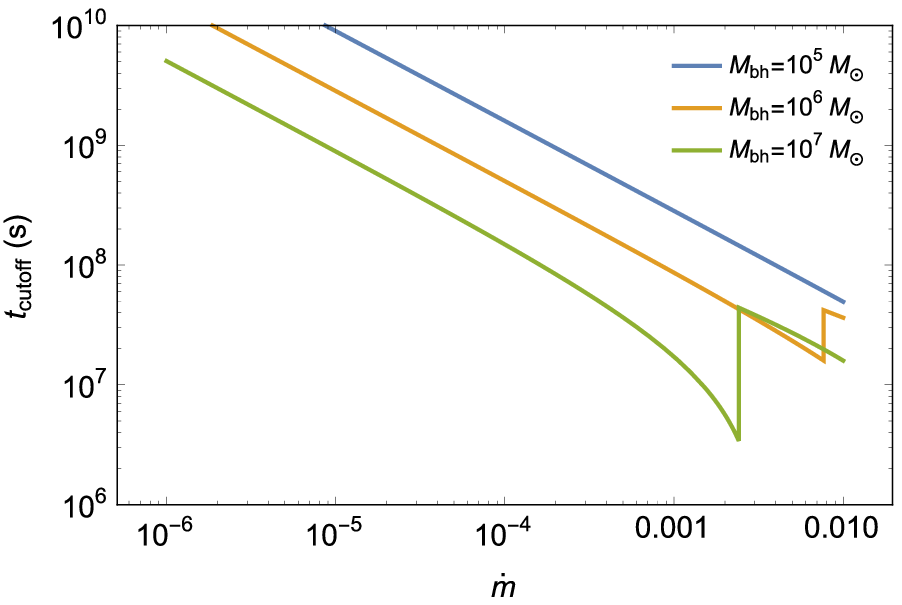}\\
\includegraphics[width=9cm]{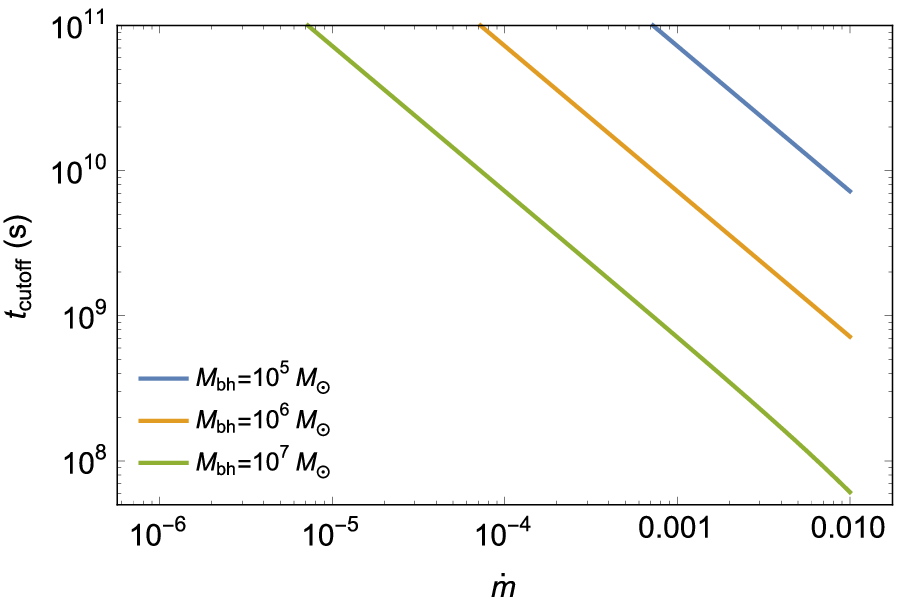}
\caption{Cut-off time versus accretion rate $\dot{m}$ (normalized to Eddington) for various SMBH masses with $\lambda=0.5$ (top panel), $\lambda=1$ (middle panel) and $\lambda=1.5$ (bottom panel). We assume a solar-type star and take $\alpha=0.01$, $\eta=0.1$ (see Section \ref{interaction} for analytic expressions of $t\sub{cutoff}$).}
\label{fig3}
\end{figure}

In order to understand the dependence of the cut-off time on the SMBH mass and also on the value of $\lambda$, we plot $t\sub{cutoff}$ versus $\dot{m}$ for multiple SMBH masses in Fig. \ref{fig3}. The different panels show different values of $\lambda=0.5, 1, 1.5$ (top, middle and bottom panels, respectively). As mentioned in Section \ref{luminosityratio}, for $\lambda=1.5$ it will be impossible to identify the cut-off time  and Fig. \ref{fig3} shows that $t\sub{cutoff}$ exceeds a century for these low-density accretion discs. Hence even the absence of a cut-off time can provide us with a lower limit for $\lambda$.

%%%%%%%%%%%%%%%%%%%%%%%%%%%%%%%%%%%
%%%%%%%%%%%%%%%%%%%%%%%%%%%%%%%%%%%

%%%%%%%%%%%%%%%%%%%%%%%%%%%%%%%%%%%
%%%%%%%%%%%%%%%%%%%%%%%%%%%%%%%%%%%

\section{Applications to Observations} \label{Applications}

As mentioned above, in the presence of a pre-existing accretion disc, the tidal disruption flare should drop in luminosity at the cut-off time (provided the flare luminosity is larger than the luminosity of the quiescent AGN at this time). Hence we can attempt to identify $t\sub{cutoff}$ from TDE observations and use it to constrain the properties of a potential pre-existing accretion disc . We apply this theory to three observed TDE candidates, {\it Swift} J1644, {\it Swift} J2058 and ASASSN-14li.

\subsection {{\it Swift} J1644 and {\it Swift} J2058}

\begin{figure}
\includegraphics[width=9cm]{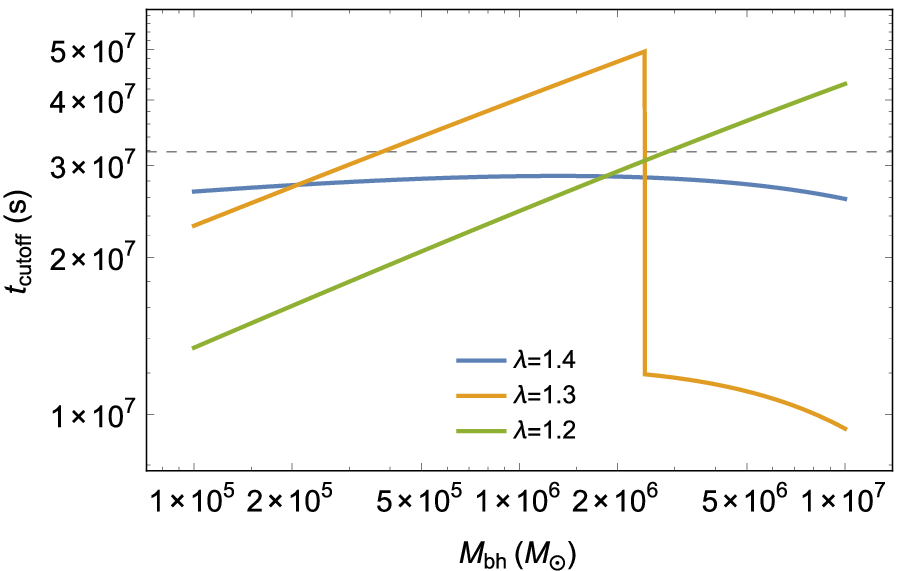}\\
\includegraphics[width=9cm]{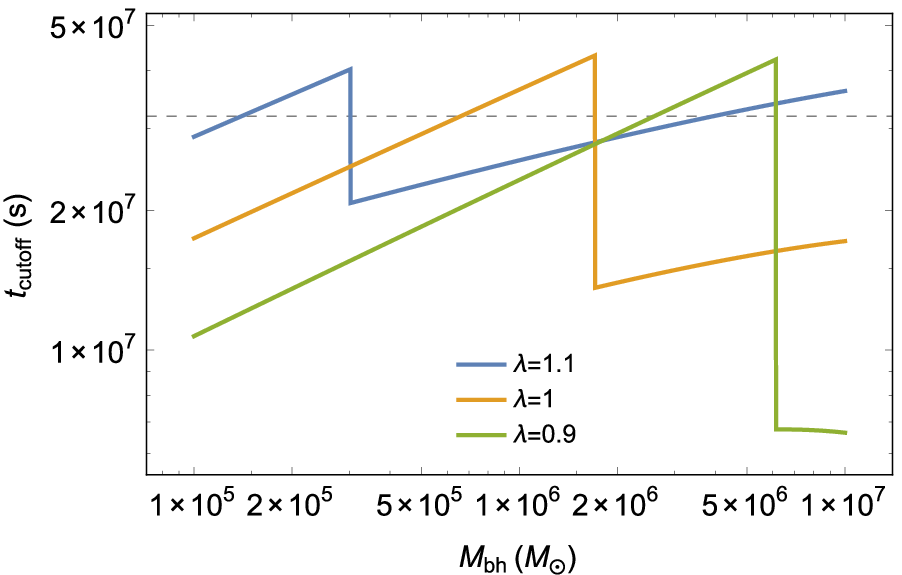}
\caption{Cut-off time versus SMBH mass for multiple values of $\lambda$. We assume a solar-type star, taking $\alpha=0.01$ (top panel) and $\alpha=0.1$ (bottom panel), fixing $\eta=0.1$ using $\dot{m}$ from equation (\ref{mdot}). The dashed line shows the observed cut-off time of TDE {\it Swift} J1644 (e.g. \citealp{burrows2011}). Given that $M_{\rm bh}$ is constrained to be $\sim 10^5-10^7 M_{\odot}$ and other values of $\lambda$ do now match the observed cut-off time for this TDE, we constrain $\lambda \sim 1.2$--$1.4$ for $\alpha=0.01$ (top panel) and $\lambda \sim 0.9$--$1.1$ for $\alpha=0.1$ (bottom panel).}
\label{fig1644}
\end{figure}

The {\it Swift} J1644 TDE candidate was initially thought to be a Gamma-ray burst (GRB). However, the long lasting X-ray flare associated with this event led to the current consensus of it being a TDE candidate (\citealp[e.g.,][]{bloom2011,burrows2011}). {\it Swift} J1644 is likely to be the first example of a jetted TDE (\citealp{giannios2011}). The X-ray emission lasted for $\sim 370$ d (in the host galaxy frame) after which the luminosity (in X-ray) dropped by two orders of magnitude to a constant value of $L\sub{x}\sim 5\times 10^{42}$ erg s$^{-1}$ and has been observed for more than 4 yr now (\citealp{levan2016}), this value is also consistent with the upper limits obtained from the pre-outburst measurements of {\it Swift} J1644 (\citealp{levan2011}). Assuming an $\sim 30$\% of the total power is emitted in X-rays, the bolometric luminosity would be $L\sim3 L\sub{x}\approx 10^{43}$ erg s$^{-1}$. We can interpret this luminosity as the quiescent accretion from the pre-existing disc on to the SMBH and take $t\sub{cutoff}$ to be the time at which the drop in X-ray luminosity was observed. The mass of the SMBH for this event is constrained within $\sim 10^5-10^7 M_{\odot}$ (\citealp{tchek2014}). Using this, we can calculate
\beq 
\dot{m}=\frac{L}{L\sub{edd}} \approx \frac{10^{43}}{10^{38}M\sub{bh}},
\label{mdot}
\eeq
where $M\sub{bh}$, $\alpha, \eta$ and $\lambda$ are the unknown parameters. Assuming a solar-type star was disrupted and fixing $\eta=0.1$, we plot $t\sub{cutoff}$ versus $M\sub{bh}$ for multiple values of $\lambda$ and $\alpha$ in Fig. \ref{fig1644}. The dashed horizontal line shows the observed cut-off time of the flare. If this cut-off occurred due to the interaction of the BDS with the pre-existing accretion disc, we can use our model to infer values of the density profile of this disc $\lambda$.
From Fig. \ref{fig1644}, we find $\lambda$ is constrained between $\sim1.2$ and $1.4$ if $\alpha=0.01$ and between $0.9$ and $1.1$ if $\alpha=0.1$. 

%%%%%%%%%%%%%%%%%%%%%%%%%%%%%%%%%%%
%%%%%%%%%%%%%%%%%%%%%%%%%%%%%%%%%%%

{\it Swift} J2058 is another TDE candidate with many similar properties to that of {\it Swift} J1644 (\citealp{cenko2012}). The flare had a sharp cut-off at $\sim200$ d, after which the luminosity (in X-rays) dropped below $8.4 \times 10^{42}$ erg s$^{-1}$ (\citealp{pasham2015}), these authors also found the SMBH mass to be constrained between $10^4$ and $10^6 M_{\odot}$. Using this upper limit on the quiescent luminosity and equation (\ref{mdot}), we find $\dot{m}\lesssim 0.3/M_6$ (assuming a pre-existing accretion disc is present). We can repeat the procedure done for {\it Swift} J1644 to find some limits on $\lambda$ (we cannot constrain $\lambda$ since a quiescent luminosity was not observed, so the value of $\dot{m}$ is not known). If $\alpha=0.01$ we do not find any limit on $\lambda$. If $\alpha=0.1$ we find $\lambda\lesssim 1.2$ for $M\sub{bh}=10^5 M_{\odot}$ and $\lambda\lesssim 1$ for $M\sub{bh}=10^6 M_{\odot}$.

\subsection{ASASSN-14li}\label{sec14li}

The flare of ASASSN-14li event shows many similar characteristics to those of other TDE candidates, such as broad hydrogen and helium lines and strong blue continuum emission (\citealp{holoien2016}). Recent X-ray observations (\citealp{brown2016}) show this flare continuing for more than 600 d (hence $t\sub{cutoff} > 600$ d). The host galaxy of ASASSN-14li shows evidence of being a weak AGN, with an estimated lower limit of $\sim 2\times 10^{41}$ erg s$^{-1}$ on its quiescent emission (\citealp{prieto2016}). Using equation (\ref{mdot}), we obtain $\dot{m}\gtrsim 2000/M\sub{bh}$, where the SMBH mass is constrained between $10^6$ and $10^7 M_{\odot}$ (e.g, \citealp{brown2016}). From these limits of $\dot{m}$ and $t\sub{cutoff}$, we can possibly rule out some accretion disc profiles using our model, as we describe below.

\begin{figure}
\includegraphics[width=9cm]{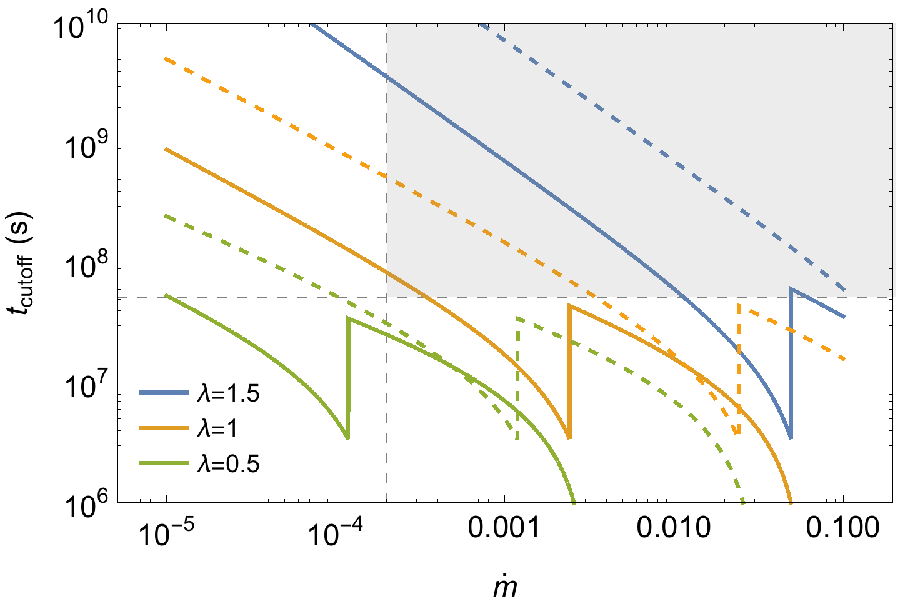}\\
\includegraphics[width=9cm]{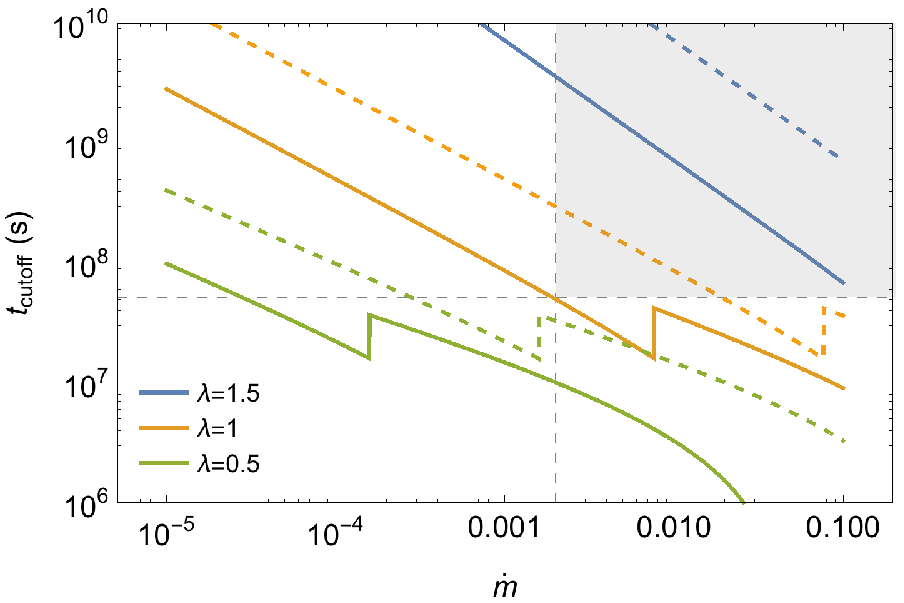}
\caption{Cut-off time versus accretion rate $\dot{m}$ for multiple values of $\lambda$ with $M\sub{bh}=10^6 M_{\odot}$ (bottom panel) and $M\sub{bh}=10^7 M_{\odot}$ (top panel). We assume a solar-type star, taking $\alpha=0.01$ (solid lines) and $\alpha=0.1$ (dashed lines), fixing $\eta=0.1$. Dashed vertical and horizontal lines represent the lower limits on $\dot{m}$ and $t\sub{cutoff}$ respectively and are inferred from observations of ASASSN-14li (\citealp{brown2016}; \citealp{prieto2016}). These inferred values restrict solutions to the shaded quadrant. }
\label{14li}
\end{figure}

In Fig. \ref{14li} we plot $t\sub{cutoff}$ versus $\dot{m}$ for multiple values of $\lambda$, $\alpha$ and $M\sub{bh}$. The dashed vertical and horizontal lines represent the lower limits on $\dot{m}$ and $t\sub{cutoff}$, respectively. Hence from observational constraints, only solutions in the shaded quadrant of each plot are valid. If $M\sub{bh}=10^6 M_{\odot}$, we find $\lambda\gtrsim 1$ for $\alpha=0.01$ and $\lambda\gtrsim 0.7$ for $\alpha=0.1$. If $M\sub{bh}=10^7 M_{\odot}$, we find $\lambda\gtrsim 1$ for $\alpha=0.01$ and we do not obtain any interesting limit on $\lambda$ if $\alpha=0.1$.

For the TDEs considered in this section, their light curve cut-offs occur at or after a few hundred days since disruption. Using equation (\ref{r}), we estimate that for a fiducial SMBH mass of $M\sub{bh}=10^6 M_{\odot}$ the distance at which we probe the pre-existing accretion disc profile is $\gae 10^{-3}$ pc. Therefore, TDE observations are able to probe the gas properties very close to the SMBH, which is an important yet difficult region to probe with other methods.

%%%%%%%%%%%%%%%%%%%%%%%%%%%%%%%%%%%
%%%%%%%%%%%%%%%%%%%%%%%%%%%%%%%%%%%

\section{Discussion and Conclusions}\label{discussion}

In this paper, we have analytically explored how TDEs are affected by a pre-existing accretion disc. In particular, we find that not all the disrupted stellar material falls into the SMBH. The fallback of stellar material can be stalled due to its interaction with the pre-existing accretion disc. This yields an abrupt cut-off in the light curve of TDEs and this cut-off time can be used to obtain valuable information on the pre-existing accretion disc density. Our framework allows us to use TDE light curves as powerful tools in diagnosing the otherwise elusive gas properties close to the SMBHs.

We have applied our theoretical framework to TDEs in which an abrupt cut-off has been observed: {\it Swift} J1644 and {\it Swift} J2508. In the first case, the sharp cut-off observed in the X-ray light curve settles to a constant luminosity, indicative of a low-luminosity AGN activity (\citealp{levan2016}). This strongly implies the presence of a pre-existing accretion disc around the SMBH for this object. These observations allow us to constrain the pre-existing accretion disc density to drop as $\propto R^{-(0.9\textendash1.4)}$ for a disc viscosity parameter of $\alpha\sim0.01$--$0.1$. In the second case, the sharp cut-off is also observed in the TDE X-ray light curve; however, only an upper limit is available for the emission after the sharp cut-off (\citealp{pasham2015}). This allows us to constrain the pre-existing accretion disc density profile to be roughly $\propto R^{-1}$ or shallower for $\alpha \sim 0.1$. Although these two cases pertain to TDEs accompanied by a jet, our theoretical framework can also be applied to TDEs without a jet. For example, in the case of ASASSN-14li, the host galaxy shows evidence of weak AGN activity (\citealp{prieto2016}), which could indicate the presence of  a pre-existing accretion disc. Even though this TDE does not display an abrupt cut-off in its light curve, a meaningful constrain can still be obtained with the absence of a cut-off. We find that the pre-existing accretion disc profile in this case must be roughly $\propto R^{-1}$ or steeper for $\alpha\sim0.01$, since shallower profiles would have already prevented the stellar material from reaching the SMBH, in which case the abrupt cut-off should have already been detected.

Radio observations of TDEs provide an independent method of constraining the gas density close to the SMBH. This is done by modelling the radio emission assuming that it originates in a shock, which is produced as a TDE outflow interacts with the gas surrounding the SMBH. For example, for the case of {\it Swift} J1644, \cite{berger2012} estimate a density profile of $\lambda \sim 1.5$ at a distance of $\sim 0.1$ pc from the SMBH (albeit with a steeper profile for $\lae 0.1$ pc). A density profile of $\lambda \sim 2.5$ is estimated for ASASSN-14li on a scale of $\sim 0.01$ pc \citep{alexander2016}. Although our findings above are roughly consistent with these profiles, our method probes the density profile at much smaller scales, of the order of $10^{-3}$ pc. Densities inferred at larger scales might reflect densities close to (or beyond) the Bondi radius.

We now briefly investigate the detectability of the emission which could arise from the shocked BDS and show that it would be observationally unimportant. To estimate the luminosity of the shocked BDS, we need the mass of the shocked material $M\sub{sh}$, which can be found by integrating $\dot{M}_{\rm fb}(t)$ in equation (\ref{Lc}) from $t\sub{cutoff}$ to infinity. From the shock jump conditions, the internal energy of this shocked material is $\approx M\sub{sh}v^2\sub{sh}$, where $v\sub{sh}$ is of the order of the sound speed ($c\sub{s}\approx1.6\times 10^6$ cm/s) at the cut-off time. This internal energy provides an upper limit for the emission of the shocked fluid. Hence  the observed luminosity should be less than $\approx M\sub{sh}v^2\sub{sh}/t\sub{cutoff}\sim 10^{35}$ erg/s, which would be quite difficult to detect.

After the BDS is stalled, it is reasonable to expect that the stalled material will accrete on to the SMBH at a much larger time given by the accretion time-scale. If the accretion of this stalled material is powerful enough, it could lead to a rise in the quiescent luminosity which would last also on the order of the accretion time-scale. From the cut-off time, we can obtain a cut-off distance using equation ({\ref{r}}) and use it to estimate the accretion time-scale ($\sim R/\alpha v_{\rm k}$). For typical parameters of $r_*, m_*, M_6, \mu$ and $\lambda$ of order unity, the accretion time-scale is $\sim 10$ yr, and the corresponding accretion rate of the stalled material is $\sim 0.01\dot{M}\sub{edd}$. Therefore, if the quiescent accretion rate of the AGN is lower than this value, the accretion of the stalled material could lead to a flare that would last $\sim 10$ yr. These calculations assume the BDS was stalled by shocks (see equation \ref{tshock}). On a more speculative note, this could explain the recent observations of a decade-long TDE flare presented in \cite{lin2017}.

So far, it seems that tidal disruption flares with jetted emission have produced an easily observable cut-off in their light curves. In addition to the two {\it Swift} candidates mentioned in this paper, {\it Swift} J1112 is another TDE candidate in which a cut-off may also have been observed (\citealp{brown2015}). This cut-off has previously been attributed to the accretion rate transitioning from super- to sub-Eddington values (e.g.,\,\citealp{decolle2012,zauderer2013,tchek2014,shen2014}). However, if this cut-off did occur due to the presence of a pre-existing disc, as proposed in this work, it would be more easily detectable in a jetted flare because its emission would be brighter due to beaming effects (see Section \ref{luminosityratio}). If this is the case, it is worth investigating the correlation between observed jetted TDE flares and AGN activity, since numerical studies show that a pre-existing disc may be required to produce a jet during TDEs (\citealp{tchek2014,kelley2014}).

In order to develop a complete analytic model, some simplifications have been made. We assume a complete stellar disruption and a penetration parameter of $\beta = r\sub{t}/r\sub{p}=1$. We further assume a spherically symmetric density distribution for the disc in order to mimic a thick pre-existing accretion disc, whereas in reality the disc will have some poloidal density structure. Also, our theoretical analysis is performed only at the apocentre passage of the BDS. In addition, we do not consider magnetic fields in the pre-existing disc nor in the BDS, but numerical studies suggest that including these fields may shorten the cut-off time. Recent simulations by \cite{guillochon2016} show that the magnetic pressure in the BDS might cause the stream to break self-gravity faster than at the time of recombination. This makes the stream more susceptible to interactions with the disc, thereby shortening the light curve cut-off time. Magnetic fields within the disc might amplify the drag force on the BDS and enhance the instabilities across the disc-stream interface, thereby stalling the BDS more efficiently (\citealp{kelley2014}). Hence the exclusion of magnetic fields in our work gives an upper limit on the cut-off times. These issues remain to be investigated through numerical simulations which follow the entire stream, consider magnetic fields in the stream and pre-existing accretion disc and also consider the accretion disc structure. Such simulations will also allow us to determine the slope of the light curve after $t\sub{cutoff}$, which cannot be predicted by our current model.

We encourage long-term observations of TDE candidates to see if an abrupt cut-off is observed in their light curves, which can be caused by a pre-existing accretion disc. Also, identifying the quiescent luminosity that the TDE light curve settles to would indicate the level at which a possible low-luminosity AGN radiates. These two pieces of information, along with a constrain on the mass of the SMBH, would allow us to determine the gas density profile well-below the Bondi radius. Even the absence of a TDE light curve cut-off yields important constrains on the gas density profile (as demonstrated in the case of ASASSN-14li). Finally, long-term observations may also be able to detect the late time emission associated with the accretion of the stalled stellar material, as briefly discussed above.

\section*{Acknowledgements}

We thank Ian M. Christie, Patrick Crumley and the anonymous referee for providing useful comments on the manuscript. We acknowledge support from NASA through grant NNX16AB32G issued through the Astrophysics Theory Program. We also acknowledge support from the Research Corporation for Science Advancement's Scialog program. 

\bibliographystyle{mn2e}
\bibliography{references} 

\begin{appendix}
\section{Conditions for a shock in the BDS}\label{shockexist}

The temperature evolution of a tidally disrupted debris stream (for a solar-type star) has been studied by \cite{kasen2010}.
Initially, the debris stream evolves adiabatically until recombination starts when $T=T\sub{rec}\sim10^4$ K,
at which point the volume of the stream is $V_{\rm rec}$. During recombination, the temperature of the star remains roughly constant until its volume has expanded to $V' \sim (15)^3 V_{\rm rec}$ \citep{kasen2010}. After this point, the stream continues to evolve adiabatically. So the temperature of the stream ($T\sub{s}$) as a function of volume is given by
\begin{subnumcases}{T\sub{s}\approx }
T_*\left(\frac{V_*}{V}\right)^{\frac{2}{3}} &$V_*\le V\le V\sub{rec}$\\
T\sub{rec}& $V\sub{rec}\le V\le V'$\\
T\sub{rec}\left(\frac{V'}{V}\right)^{\frac{2}{3}}& $V\ge V'$ \label{Ts}
\end{subnumcases}
where $V_*$ and $T_*$ are the initial volume and temperature of the star, and
\beq
V=\pi\left(\frac{h}{2}\right)^2r
\eeq
is the volume of the stream. Using equations (\ref{h}) and (\ref{r}), we can obtain how $T\sub{s}$ evolves with time. For the relevant time-scales discussed in the main text, we need only (\ref{Ts}) to describe the temperature of the BDS. We find
\beq
T\sub{s}\approx\frac{(15)^2\pi^{\frac{4}{9}}2^{\frac{1}{6}}G^{\frac{7}{9}}m\sub{p}}{3^{\frac{2}{3}}c\sub{s}^{\frac{4}{3}}k\sub{b}}\,R_*M_*M\sub{bh}^{-\frac{2}{9}}t^{-\frac{16}{9}},
\eeq
where $k\sub{b}$ is the Boltzmann constant. 

The temperature of the shocked debris fluid ($T\sub{sh}$) can be obtained using the shock jump conditions 
\beq
T\sub{sh}\approx\frac{5}{16}\frac{m\sub{p}}{\gamma k\sub{b}}v\sub{sh}^2+\frac{7}{8}T\sub{s}\approx 2\times 10^{-9}v\sub{sh}^2+\frac{7}{8}T\sub{s},
\label{Tsh}
\eeq
where $\gamma=5/3$ and $v\sub{sh}$ is given in equation (\ref{vsh}).
In order for a shock to exist (at $t\sub{shock}$), we require the Mach number, $v\sub{sh}/c\sub{sound}\ge 1$. From the Rankine--Hugoniot conditions we find this is equivalent to saying
$T\sub{sh}/T\sub{s}\ge1$. Evaluating this temperature ratio at $t\sub{shock}$ we find
\beq
\begin{split}
\frac{T\sub{sh}}{T\sub{s}}(t\sub{shock})=\frac{c\sub{s}^{\frac{10}{3}}}{\gamma R_*}\Bigg(&\frac{3^{-5}5^{3}c^{-16}}{2^{\frac{219}{2}}\pi^{12}G^{7}}\,\bigg(\frac{\mu\sigma\sub{T}}{m\sub{p}}\bigg)^8\frac{M\sub{bh}^{2}}{M_*}\\ &\times\left(\frac{3^7\,2^{\frac{49}{2}}c^{16}M_*^3}{\pi^4\,5^5\,G^3M\sub{bh}^6}\right)^{\lambda}\Bigg)^{\frac{1}{3(3-\lambda)}}+\frac{7}{8}.
\end{split}
\eeq
From this equation we can, for instance, find a constraint on the SMBH mass for which a shock exists, and it is given by
\beq
\begin{split}
M\sub{bh}\le  \Bigg(&\frac{3^{-5}\,5^{3}\,c\sub{s}^{30}c^{-16}\gamma^{-9}}{2^{\frac{165}{2}}\pi^{12}\,G^{7}R_*^9\,M_*}\,\bigg(\frac{\mu\sigma\sub{T}}{m\sub{p}}\bigg)^8\\ &\times\left(\frac{3^7\,2^{\frac{31}{2}}c^{16}\gamma^3R_*^3M_*^3}{\pi^4\,5^5\,G^3\,c\sub{s}^{10}}\right)^{\lambda}\Bigg)^{\frac{1}{2(3\lambda-1)}}.
\end{split}
\eeq
This expression allows us to analytically determine the SMBH mass at which the stalling mechanism of the BDS transitions from shock crossing to momentum imparted by the disc (this transition is seen as a discontinuous jump in Fig. \ref{fig1} and Fig. \ref{fig1644}).

\end{appendix}

\end{document}